\documentclass[aps,amsmath,amssymb,reprint,groupedaddress,floatfix]{revtex4-1}

\usepackage{graphicx}
\usepackage{floatrow}
\usepackage[caption=false,font={large}]{subfig}
\usepackage{bm}

\usepackage{amsmath}
\usepackage{amssymb}
\usepackage{color}
\usepackage{times}
\usepackage{float}
\usepackage[sort&compress]{natbib}

\usepackage[colorinlistoftodos,prependcaption,textsize=small]{todonotes}

\usepackage{hyperref}

\hypersetup{
  colorlinks=true,
citecolor=black,
linkcolor=black,
urlcolor=black
  }

\usepackage[T1]{fontenc}
\usepackage[utf8]{inputenc}

\begin{document}

\title{Drifted escape from the finite interval}
\author{Karol Capa{\l}a}
\email{karol@th.if.uj.edu.pl}

\author{Bart{\l}omiej Dybiec}
\email{bartlomiej.dybiec@uj.edu.pl}

\affiliation{Institute of Theoretical Physics, Department of Statistical Physics, Jagiellonian University, \L{}ojasiewicza 11, 30-348 Krak\'ow, Poland}


\date{\today}

\begin{abstract}
Properties of the noise-driven escape kinetics are mainly determined by the stochastic component of the system dynamics.
Nevertheless, the escape dynamics is  also sensitive to deterministic forces.
Here, we are exploring properties of the overdamped drifted escape from finite intervals under the action of symmetric $\alpha$-stable noises.
We show that the properly rescaled mean first passage time follows the universal pattern as a function of the generalized P\'ecklet number, which can be used to efficiently discriminate between domains where drift or random force dominate.
Stochastic driving of the $\alpha$-stable type is capable of diminishing the significance of the drift in the regime when the drift prevails.

\end{abstract}


%
\maketitle

\setlength{\tabcolsep}{0pt}

%
%

%
%
\section{Introduction}

Noise is considered as the effective approximation of the complicated interactions of the test particle with the environment.
In the simplest situations, it is assumed that the noise is Gaussian and white, i.e., the noise kicks are bounded (characterized by a finite variance) and independent.
Nevertheless, this assumption can be violated.
In particular, the noise can still be of the white type, but increments do not need to follow the Gaussian distribution.
The natural generalization of the Gaussian density is the $\alpha$-stable distribution.
$\alpha$-stable densities have analogous algebraic properties \cite{samorodnitsky1994,janicki1994} like the Gaussian density.
Due to the generalized central limit theorem \cite{samorodnitsky1994}, they are limiting distributions for sums of independent and identically distributed random variables characterized by diverging variance.
Therefore, the natural generalization of the GWN is provided by $\alpha$-stable L\'evy type noise, which is still white, but the increments follow a power-law distribution.
Non-Gaussian, heavy-tailed fluctuations have been observed in various experimental setups like climate dynamics \cite{ditlevsen1999b}, turbulent systems \cite{shlesinger1986b},   hot atomic vapors \cite{mercadier2009levyflights}, anomalous diffusion in laser cooling \cite{cohen1990,barkai2014},  gaze dynamics \cite{amor2016}, memory retrieval in humans \cite{reynolds2007}  and financial time series \cite{bouchaud1990}.
Additionally, L\'evy noise underlines two well-known stochastic models of anomalous diffusion:  L\'evy flights \cite{shlesinger1995} and L\'evy walks \cite{klafter1996,zaburdaev2015levy}.
Non-Gaussian L\'evy noises and L\'evy statistics are commonly encountered in the studies of extreme events \cite{stanley1986,mantegna2000,newman1999} or foraging strategies \cite{humphries2012,viswanathan2011physics}.
L\'evy noises, have been also applied in various theoretical considerations \cite{metzler2000,barkai2001,anh2003,brockmann2002,chechkin2006,jespersen1999,yanovsky2000,schertzer2001}.

The overdamped Langevin equation provides a common framework for the description of stochastic systems.
Numerous theoretical and experimental studies have proven that the properties of stochastic systems are determined by the noise type and deterministic forces.
On the single particle level, the model of escape kinetics from a finite interval is defined by the overdamped Langevin equation driven by L\'evy noise.
Trajectories of the underdamped motions driven by L\'evy noise are no longer continuous.
This is a natural consequence of the fact that under L\'evy driving long jumps are significantly more likely to occur than  under the action of the Gaussian white noise.
Mathematically, this property of $\alpha$-stable processes is manifested by the decomposition \cite{ditlevsen1999, imkeller2006, imkeller2006b} of $\alpha$-stable noise into a compound Poisson process describing long jumps and the Wiener part responsible for small displacements.
Discontinuity of trajectories makes the escape scenario very different \cite{ditlevsen1999,pavlyukevich2010,bier2018,capala2020levy}.
Instead of approaching the end of the interval (or any point), a particle can jump over it.
The possibility of long jumps is also responsible for emergence of the difference between first arrival and first passage for L\'evy driven systems \cite{koren2007,koren2007b,dybiec2016jpa} and for the breakdown of methods of images \cite{chechkin2003b}.

Here, we are assuming that in addition to noise, there is a constant drift, which is capable of inducing persistent motion.
Therefore, we study the interplay between the noise and the deterministic drift.
This interplay is well understood in the case of Gaussian white noise, but far less is known for L\'evy driving.
We want to assess how the possibility of long jumps affects drift-induced persistence of motion and escape kinetics.

The model under study is presented in the Sec.~\ref{sec:model} (Model).
Results of numerical simulation are given in Sec.~\ref{sec:results} (Results).
The paper is closed with the summary and conclusions (Sec.~\ref{sec:summary}).
Auxiliary material is provided in two appendices: App.~\ref{sec:gwndriving} (Gaussian white noise driving) and~App.~\ref{sec:lwndriving} (L\'evy driving).

%
%
\section{Model\label{sec:model}}
The escape of an overdamped particle from a finite interval restricted by two absorbing boundaries is a common research motif.
Such a system is described by the overdamped Langevin equation
\begin{equation}
  \frac{dx}{dt}=\sigma\xi(t),
  \label{eq:langevin-free}
\end{equation}
where $\xi(t)$ represents the stochastic force --- noise, while $\sigma$ scales its strength.
The basic measure which characterizes the escape kinetics is the mean first passage time, which is the average of first passage times --- times which are needed to leave the domain of motion for the first time.
It is very convenient to assume that the noise term in Eq.~(\ref{eq:langevin-free}) is of the Gaussian white type.
However, the formula for the MFPT is also known for more general noises --- $\alpha$-stale noises.
Therefore, we assume that $\xi(t)$ stands for the general, symmetric $\alpha$-stable noise.
The symmetric L\'evy noise $\xi(t)$ is a formal time derivative of symmetric $\alpha$-stable process $L(t)$, see \cite{janicki1994,samorodnitsky1994}.
Symmetric $\alpha$-stable process might be defined by its characteristic function
\begin{equation}
 \phi(k)=\langle \exp[i k L(t)] \rangle=\exp\left[ - t |k|^\alpha \right],
 \label{eq:fcharakt}
\end{equation}
with $\alpha$ ($\alpha \in (0,2]$) standing for the stability index.
Consequently, the increments of the symmetric $\alpha$-stable process follow the symmetric $\alpha$-stable density, which for $\alpha<2$ has the power-law asymptotics of the $|x|^{-(\alpha+1)}$ type.
For $\alpha=2$, the L\'evy noise reduces to the Gaussian white noise, because $\alpha$-stable motion reduces to the Wiener process.

The Langevin equation needs to be accompanied with appropriate boundary conditions.
Here, we are applying one of the common setups.
Absorbing boundaries are placed at $\pm L$, i.e., the motion is restricted to the $(-L,L)$ interval.
For such a setup with any initial condition, the MFPT
\begin{equation}
    \mathcal{T} =   \langle t_{\mathrm{fp}} \rangle =
     \langle \min\{t : x(0)=x_0 \;\land\; |x(t)| \geqslant L \} \rangle
     \label{eq:mfpt-definition}
\end{equation}
reads \cite{getoor1961,widom1961stable,kesten1961random,kesten1961theorem,zoia2007}
\begin{equation}
    \mathcal{T}(x_0)=\frac{(L^2-x_0^2)^{\alpha/2}}{\Gamma(1+\alpha)\sigma^\alpha}.
    \label{eq:mfpt-alpha}
\end{equation}
For more detailed information see App.~\ref{sec:lwndriving}.

The model setup described by Eq.~(\ref{eq:langevin-free}) can be generalized in multiple manners.
Here, we are assuming that in addition to the noise term there is a constant drift $\mu$.
Therefore, we are studying the overdamped motion described by the following Langevin equation
\begin{equation}
  \frac{dx}{dt}=\mu+\sigma\xi(t),
  \label{eq:langevin}
\end{equation}
on the finite interval restricted by two absorbing boundaries located at $\pm L$, i.e., $x\in (-L,L)$.
Please note that the constant drift represents the motion in a static linear potential $V(x)=-\mu x$.
The presence of the constant bias induces a preferred direction of motion which introduces a competition between isotropic stochastic driving and the deterministic motion.
The properties of drifted escape under GWN driving are discussed in~App.~\ref{sec:gwndriving}.

Using the characteristic scales of the system, one can define the set of transformations
\begin{equation}
\left\{
\begin{array}{lcl}
\tilde{x} & = &  x/L, \\
\tilde{t} & = & t \mu/L
\end{array}
\right.,
\label{eq:transformation}
\end{equation}
to rewrite Eq.~(\ref{eq:langevin}) in the dimensionless form
\begin{equation}
    \frac{d\tilde{x}}{d\tilde{t}}=1+\tilde{\sigma}\tilde{\xi}(\tilde{t}).
  \label{eq:langevinUnitless}
\end{equation}
After such a transformation, the motion is restricted to the $(-1,1)$ interval under the action of fixed to the unity drift.
The number of relevant parameters is reduced to the stability index $\alpha$ and the dimensionless noise intensity
\begin{equation}
    \tilde{\sigma}=\frac{\sigma}{\mu^{1/\alpha}L^{1-1/\alpha}}.
\end{equation}

For $x_0=0$, the formula~(\ref{eq:mfpt-alpha}) rewritten in the units defined by the transformation~(\ref{eq:transformation})  reads
\begin{equation}
    \tilde{T} = \frac{\mu}{L} T = \frac{1}{\Gamma(\alpha+1)}  \frac{\mu}{L} \left( \frac{L}{\sigma}\right)^{\alpha}.
    \label{eq:mfptdimensionless}
\end{equation}
Eq.~(\ref{eq:mfptdimensionless}) can be further shortened using the generalized P\'eclet number \cite{palyulin2014}
\begin{equation}
    \kappa = \frac{\mu L^{\alpha-1}}{\Gamma(\alpha+1) \sigma^{\alpha}}
    \label{eq:generalizedpeclet}
\end{equation}
to $\tilde{T} = \kappa.$
For $\alpha=2$ the generalized P\'eclet number $\kappa$ reduces to standard  P\'eclet number
\begin{equation}
    \mathrm{Pe}=\frac{\mu L}{2 \sigma^2},
\end{equation}
which measures the ratio between advective and diffusive transport rates.
The generalized P\'eclet number $\kappa$ can be efficiently used to examine the competition between deterministic and random motions.

In the situation when the stochastic term is much larger than the deterministic (constant) force, i.e., $\kappa \ll 1$, the drift term in Eq.~(\ref{eq:langevin}) is negligible and we can assume that MFPT will follow the solution for the (deterministic) force-free escape.
For a particle starting at the origin \cite{getoor1961,widom1961stable,kesten1961random,kesten1961theorem,zoia2007}
\begin{equation}
    \tilde{T} =  \kappa.
    \label{eq:mfptSmallKappa}
\end{equation}
For any initial position $x_0$, one obtains
\begin{equation}
    \tilde{T} =\kappa \left(1-\tilde{x}_0^2\right)^{\alpha/2}.
    \label{eq:mfptSmallKappa2}
\end{equation}
In the opposite limit of weak diffusion ($\kappa \gg 1$), i.e., when the drift is much stronger than the stochastic term, the escape is almost exclusively deterministic.
Therefore, the MFPT is given by
\begin{equation}
    \tilde{T} = \frac{\mu}{L} T =  \frac{\mu}{L} \frac{L-x_0}{\mu} = (1-\tilde{x}_0).
    \label{eq:mfptLargelKappa}
\end{equation}
For clarity of the presentation, in further considerations we will omit tildes over dimensionless units.

The interpretation of the generalized P\'eclet number $\kappa$ in the relation to stochastic and deterministic drivings seems straightforward.
Nevertheless, some special care is required if one wants to control $\kappa$ by changing the system size, since the exponent $\alpha-1$ changes the sign for $\alpha=1$, see Eq.~(\ref{eq:generalizedpeclet}).
For $\alpha>1$ increase in the interval width emphasizes significance of the drift and therefore $\kappa$ increases with $L$.
On the other hand, for $\alpha<1$ the opposite behavior is observed, i.e., $\kappa$ decreases with $L$, since the larger interval favors stochastic escapes induced by long jumps.
Finally for $\alpha=1$, the change in interval half-width cannot move the system from the drift dominating regime to the diffusion prevailing and vice versa.

%
%
\section{Results\label{sec:results}}

\begin{figure}[!h]
    \centering
    \includegraphics[angle=0,width=0.95\columnwidth]{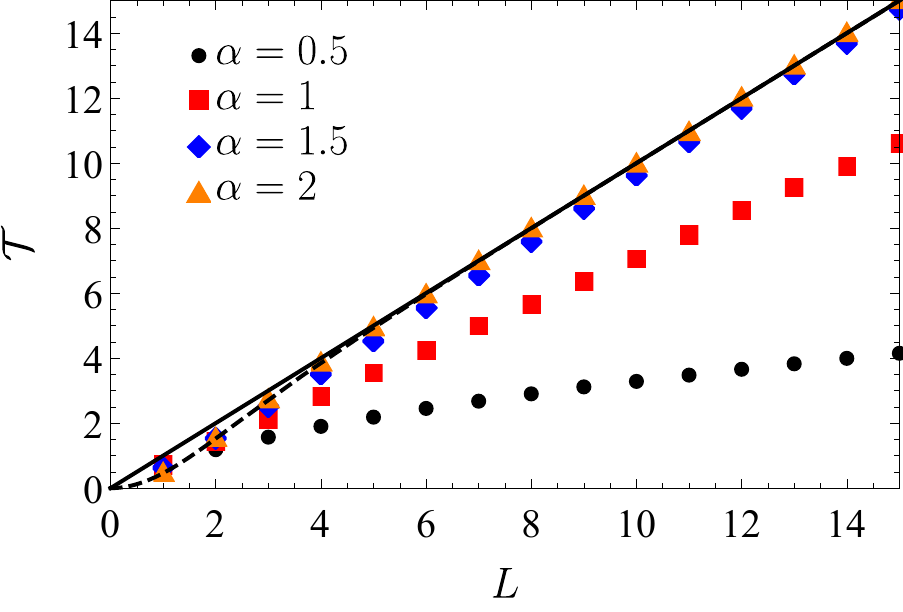}
    \caption{MFPT as a function of the interval half-width $L$ and different values of the stability index $\alpha$ for the particle starting in the middle of the interval, i.e., $x_0=0$.
    The solid line presents linear growth, i.e., $\mathcal{T}=L$, the dashed line corresponds to the analytical solution for GWN driving ($\alpha=2$) given by Eq.~(\ref{eq:mfpt-full}), while points show results of numerical simulations with different values of the stability index $\alpha$.
    The noise strength ($\sigma$) and the drift term ($\mu$) are set to unity, i.e., $\sigma=1$ and $\mu=1$.}
    \label{fig:mfptunits}
\end{figure}

\begin{figure}[!h]
    \centering
     \includegraphics[angle=0,width=0.95\columnwidth]{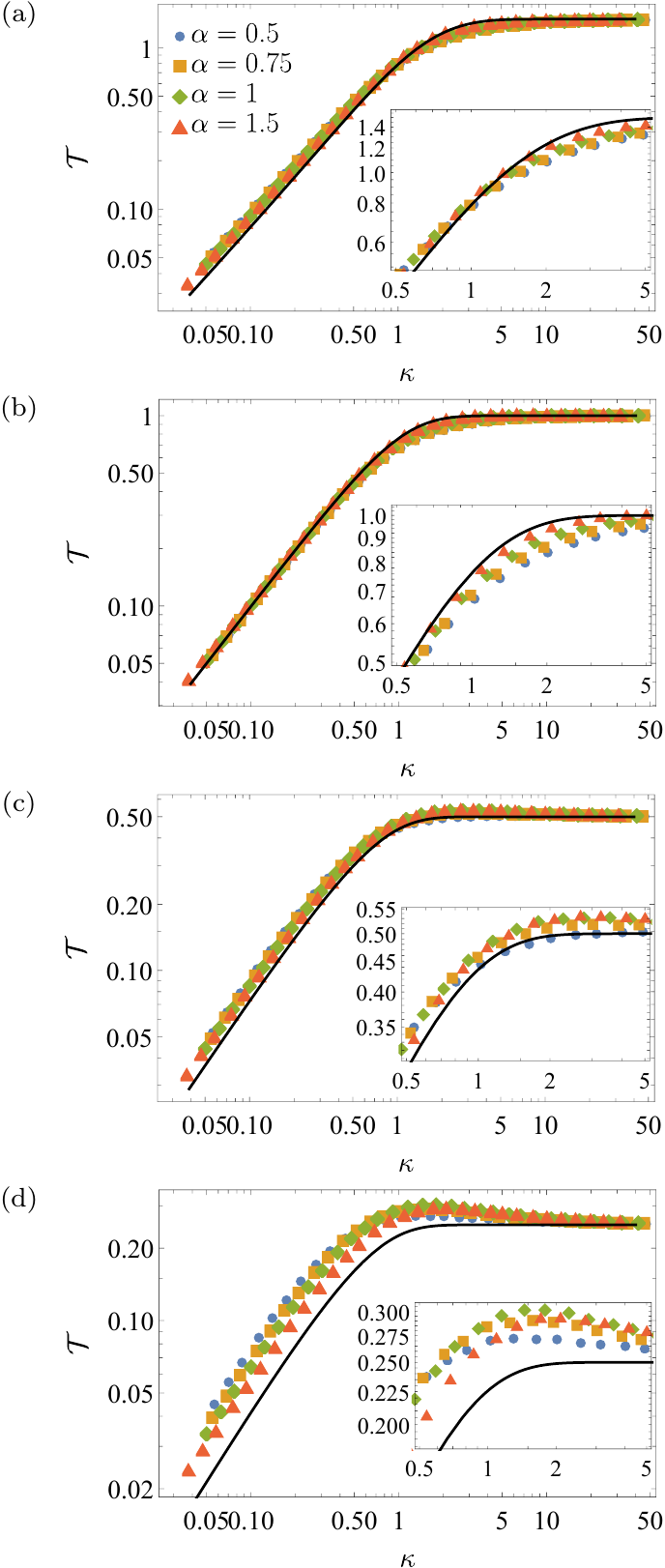}
    \caption{Dimensionless MFPT as a function of the generalized P\'ecklet number $\kappa$. Solid line corresponds to the transformed (dedimensionalized) analytical solution for GWN ($\alpha=2$) given by Eq.~(\ref{eq:mfpt-full}) while points show numerical results for different values of stability index $\alpha$. Different rows correspond to different initial positions $x_0 \in\{-0.5,0,0.5,0.75\}$ from top (panel (a)) to bottom (panel (d)) respectively.
    Insets zoom the $\kappa \simeq 1$ region.}
    \label{fig:mfptBothSides}
\end{figure}

\begin{figure}[!h]
    \centering
     \includegraphics[angle=0,width=0.95\columnwidth]{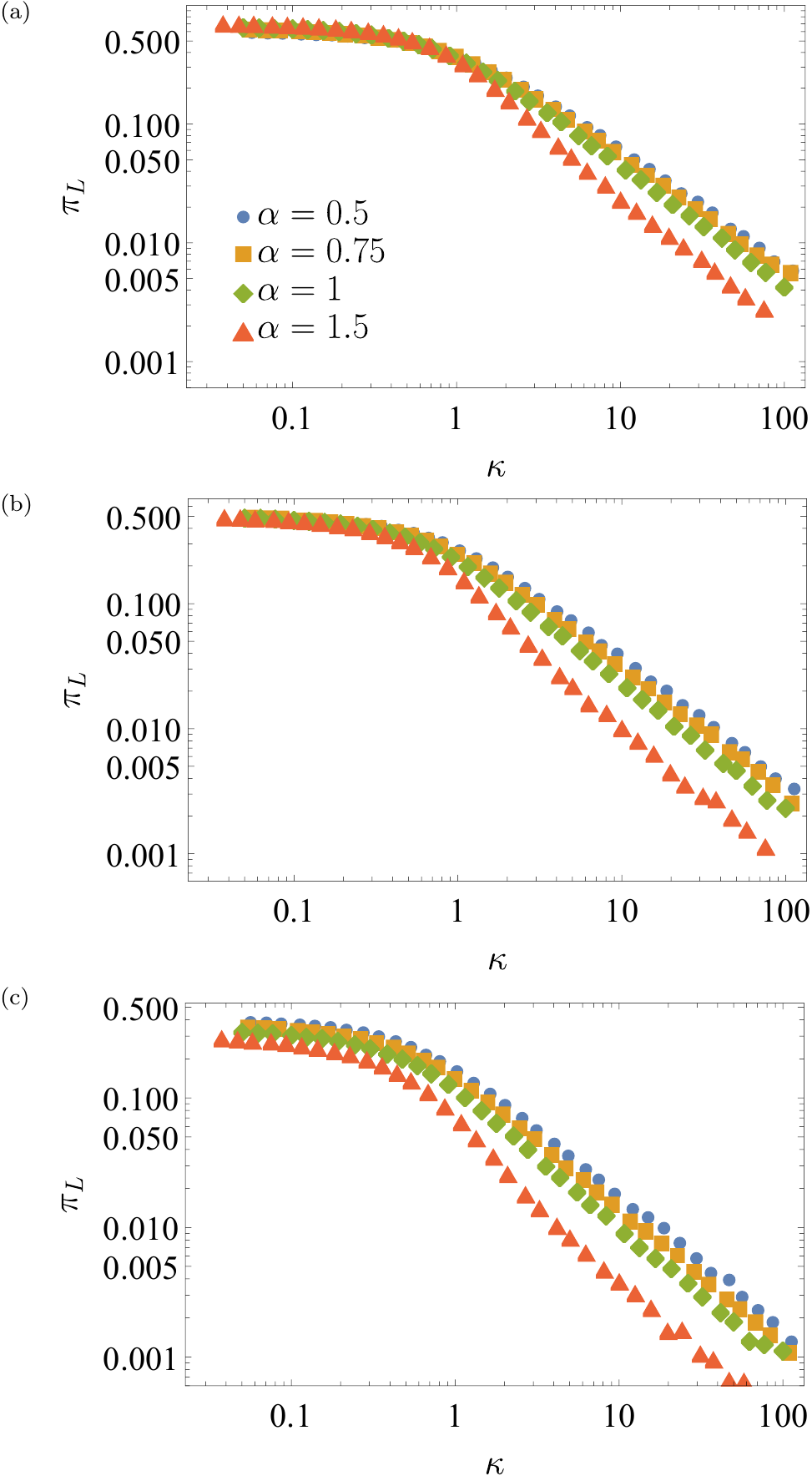}
    \caption{Probability of escaping through the left boundary, i.e, $x=-1$, as a function of the generalized P\'ecklet number $\kappa$.
    Different rows correspond to different initial positions $x_0 \in\{-0.5,0,0.5\}$ from top (panel (a)) to bottom respectively (panel (c)).}
    \label{fig:splittingBothSides}
\end{figure}

\begin{figure}[!h]
    \centering
    \includegraphics[angle=0,width=0.95\columnwidth]{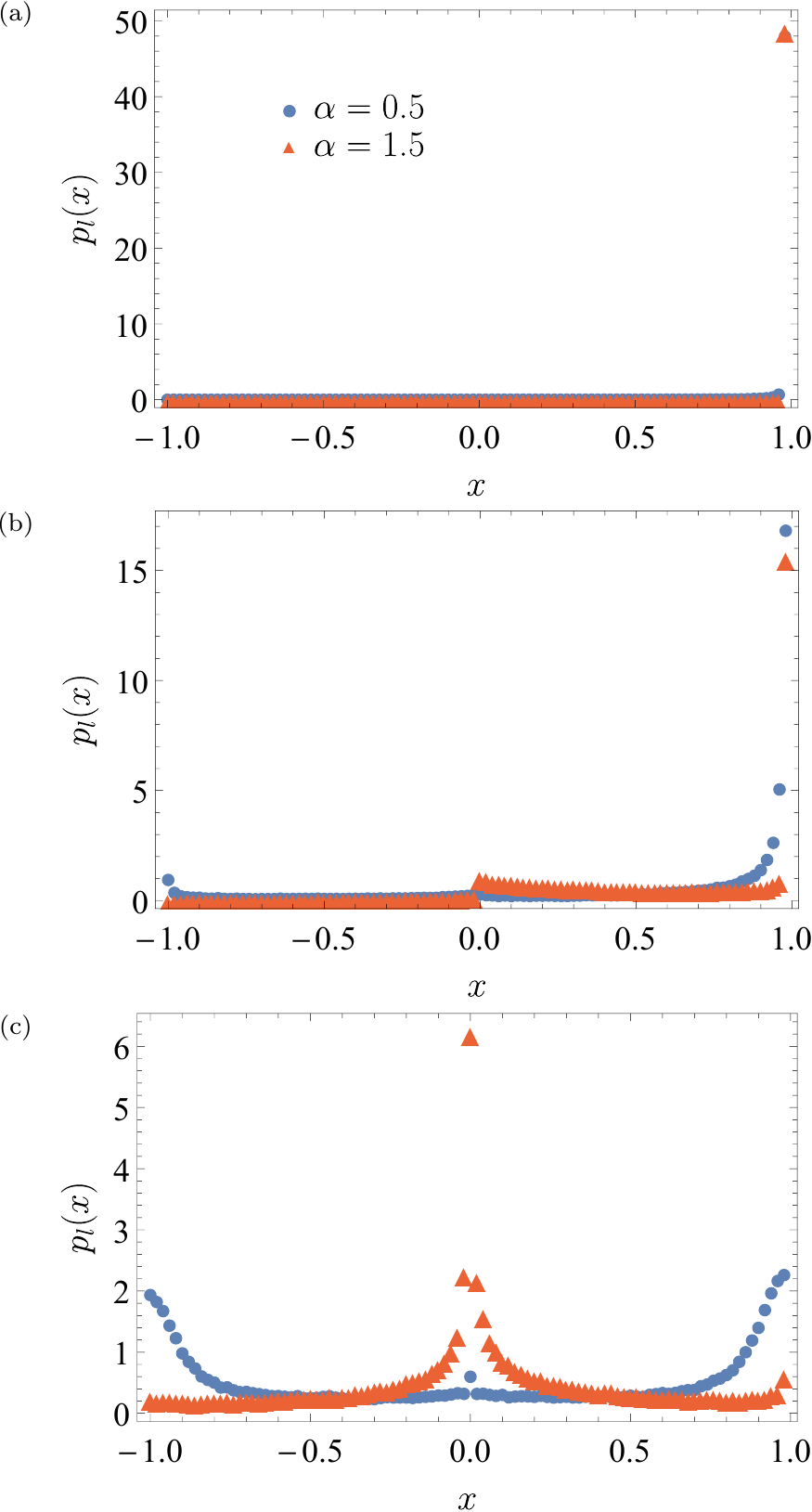}
    \caption{Distribution of last hitting points $p_l(x)$ for the particle starting in the middle of the interval, i.e., $x_0=0$ with $\kappa=47$ (top panel --- (a)), $\kappa=1$ (middle panel --- (b)) and $\kappa=0.57$ (bottom panel --- (c)).
    Various points represent different values of the stability index $\alpha$:  $\alpha=0.5$ (blue bullets) and $\alpha=1.5$ (red triangles).
    }
    \label{fig:LHP}
\end{figure}

The model described by Eq.~(\ref{eq:langevin}) was studied numerically.
Trajectories $x(t)$ have been discretized  with the help of the Euler-Maruyama scheme \cite{higham2001algorithmic,mannella2002}
\begin{equation}
    x(t+\Delta t) = x(t) + \sigma \xi_i \Delta t^{1/\alpha},
\end{equation}
where $\xi_i$ is the sequence of independent identically distributed random variables following the symmetric $\alpha$-stable density \cite{chambers1976,weron1995,weron1996}.
From the ensemble of trajectories, the MFPT has been approximated as the mean value of the recorded first passage times, see Eq.~(\ref{eq:mfpt-definition}).
In addition to the MFPT (Figs.~\ref{fig:mfptunits} and~\ref{fig:mfptBothSides}) we explore splitting probability $\pi_L$ (Fig.~\ref{fig:splittingBothSides}) and last hitting point density $p_l(x)$ (Fig.~\ref{fig:LHP}).

To fully emphasize the role of dimensionless units in Fig.~\ref{fig:mfptunits} we show the MFPT, in the dimensional units, as a function of the interval half-width $L$ and different values of the stability index $\alpha$ with the scale parameter $\sigma$ and the drift $\mu$ set to the unity and $x_0=0$.
The solid line visible in Fig.~\ref{fig:mfptunits} depicts the linear growth ($\mathcal{T}=L$), while the dashed line shows the theoretical dependence of the MFPT for $\alpha=2$, see Eq.~(\ref{eq:mfpt-full}).
For the increasing interval half-width $L$, the drift becomes the main factor facilitating the escape kinetics.
Therefore, one can expect that MFPT grows linearly with $L$.
For the Gaussian white noise, the linear asymptotic of the MFPT is perfectly visible.
For small $L$ with $\alpha=2$ deviations from the linear scaling are visible, however, the results of numerical simulations perfectly follow the theoretical prediction given by Eq.~(\ref{eq:mfpt-full}).
The linear asymptotic of the MFPT is also recorded for decreasing values of $\alpha$, but at this time the transient regime increases with the decrease of the stability index $\alpha$.
Fig.~\ref{fig:mfptunits} depicted in the dimensional units should be contrasted with Fig.~\ref{fig:mfptBothSides} using the dimensionless units, see below.

Main properties of escape kinetics from a finite interval restricted by two absorbing boundaries are determined by the generalized P\'ecklet number $\kappa$.
Consequently, the behavior of MFPT can be divided into three distinct regimes based on the $\kappa$ parameter.
Two of them, namely $\kappa\ll1$ (diffusion prevails) and $\kappa\gg1$ (drift dominates), were already mentioned in the previous section.
The dimensionless MFPT in these two regimes is given by Eq.~(\ref{eq:mfptSmallKappa}) ($\kappa \ll 1$) and Eq.~(\ref{eq:mfptLargelKappa}) ($\kappa \gg 1$) respectively.
The third region is a transitive one as it corresponds to $\kappa \simeq 1$.
In this case, both the drift and the noise are of the similar order and the analytical approximation based on drift-free solution does not fully work.
Interestingly, the most unexpected behavior can be found there, as we will discuss shortly.

Fig.~\ref{fig:mfptBothSides} shows the dimensionless MFPT as a function of $\kappa$ for different values of stability index $\alpha$ with various initial conditions $x_0$ ($x_0 \in \{-0.5,0,0.5,0.75\}$ from top to bottom).
Solid lines correspond to the transformed (dedimensionalized) analytical solution for $\alpha=2$, see Eq.~(\ref{eq:mfpt-full}).
For all shown initial positions $x_0$, the three aforementioned regimes are well visible.
When the drift dominates ($\kappa \gg 1$), dimensionless MFPT has (practically) constant value defined by a deterministic escape.
Moreover, as predicted by Eq~(\ref{eq:mfptLargelKappa}), for large $\kappa$, MFPT depends only on the initial position $x_0$.
For the opposite limit of small $\kappa$ (diffusion prevails), results practically do not depend on $\alpha$ for $x_0=0$ only, which is the symmetric initial condition, see Eq.~(\ref{eq:mfptSmallKappa}).
Nevertheless, the differences between the results corresponding to different values of stability indices for $x_0=\pm 0.5$ are rather small.
Much larger disagreement visible for $x_0=0.75$ comes from $\alpha$-dependent factor $\left(1-x_0^2\right)^{\alpha/2}$,  see Eq.~(\ref{eq:mfptSmallKappa2}), whose spread for $\alpha \in (0,2]$ depends on $x_0$.
For $x_0=0.75$, it scales with the value of stability index like $0.4375^{\alpha/2}$, while for $x_0=\pm0.5$ it behaves like $0.75^{\alpha/2}$. 
Therefore, the closer to the boundary initial position is, the larger differences between various values of stability index $\alpha$ are. 
Nevertheless, for $\kappa\ll1$, the linear growth predicted by Eq.~(\ref{eq:mfptSmallKappa}) was observed for all studied initial positions.
Finally, in the regime of $\kappa \simeq 1$, MFPTs for different values of stability index $\alpha$  always differ.
In this regime, for $x_0=0$ and $x_0=-0.5$,  MFPT is a decreasing function of $\alpha$, while for $x_0=0.5$ and $x_0=0.75$ it behaves non-monotonously, see Fig.~\ref{fig:mfptBothSides}.
Importantly, for $\kappa\simeq1$ a transition between regimes dominated by drift and diffusion takes place.
This transition can occur in two ways, influencing monotonicity of the MFPT as a function of $\kappa$.
For the escape driven by GWN or for non-positive initial position $x_0$, MFPT is an increasing function of $\kappa$.
Therefore, asymptotically the slowest escape can be observed when a particle is forced fully deterministic, i.e., $\kappa\to\infty$.
However, for the escape process with the positive initial position, MFPT as the function of $\kappa$ may have maximum for $\kappa \simeq 1$, see panels (c) and (d) of Fig.~\ref{fig:mfptBothSides}.
Occurrence of this effect depends on both the stability index $\alpha$ and the initial position.
The closer to the boundary pointed by the drift the initial position is, the maximum appears for a wider range of $\alpha$ and the more pronounced it is.
For example for $x_0=0.5$, see panel (c) of Fig.~\ref{fig:mfptBothSides}, this behavior is visible for $\alpha \in \{0.75,1,1.5\}$ but not for $\alpha=0.5$ and $\alpha=2$ (GWN).
At the same time, for $x_0=0.75$, the pronounced maximum can be observed for all considered non-Gaussian cases, see panel (d) of Fig.~\ref{fig:mfptBothSides}.
The slowest escape for $\kappa\simeq1$ can be attributed to the competition between deterministic and stochastic forces as both of them play an important role in the $\kappa \simeq 1$ regime.
Drift leads to deterministic escape through the right boundary, introducing asymmetry into the system.
Small noise `kicks' can slightly disturb such escape, but long jumps are essential to understand the occurrence of the maximum of MFPT.
Due to asymmetry of the system and initial conditions, it is more likely that long jump, which does not lead to the escape, displaces particle to the position from which drift-driven escape is longer than from the one before the `kick'.
Maximum of MFPT can be observed when the slowdown in escape caused by this effect cannot be compensated by single-jump escapes.

Fingerprints of the three regimes visible for MFPT can be seen also in the splitting probability, i.e., the probability of leaving the domain of motion via a particular end of the interval.
Fig.~\ref{fig:splittingBothSides} shows the probability $\pi_L$ that a particle escapes from the system via the left boundary for three different initial conditions $x_0$ ($x_0\in \{-0.5,0, 0.5\}$ from top to bottom).
Please note that for $\mu>0$ the drift favors the motion to the right, therefore escapes via the right boundary seems to be more likely.
Despite the quantitative differences between different $\alpha$ with various $x_0$, the qualitative behavior of all curves is very similar.
For small $\kappa$, the splitting probability does not change significantly and it is very close to its drift-free value.
For $x_0=0$, this probability is equal to $0.5$, since in the absence of drift the problem is fully symmetric.
Therefore, since $\kappa \ll 1$ corresponds to a weak drift limit, the asymmetry caused by the deterministic force can be neglected without significant loss in prediction accuracy.
Similar explanation can be used for $x_0=\pm 0.5$, if one takes into account the asymmetry caused by the initial conditions.
Further increase of $\kappa$ leads to the regime where both forces in the system --- stochastic and deterministic, compete with each other.
This regime for the splitting probability, contrary to MFPT, is not limited to $\kappa\simeq 1$ only.
The probability of escaping through the barrier opposite to the direction of the drift displays power-law decay, i.e.,
\begin{equation}
    \pi_L \sim \kappa^{-a}.
    \label{eq:pi-pl}
\end{equation}
Therefore, the purely drift-driven behavior can be observed only for $\kappa \to \infty$.
The key factor to understand this behavior arises due to the competition between drift and noise, which is capable of producing heavy-tailed noise pulses.
For $\kappa \gg 1$, the drift plays an essential role in the system, because it persistently pushes particles towards one of the barriers.
Nevertheless, even for a very strong but finite drift, the escape is not instantaneous.
Rare but strong noise pulses are still capable of pushing particles towards the opposite barrier.
The competition between deterministic and random forces can explain the order of splitting probabilities $\pi_L$ in the Fig.~\ref{fig:splittingBothSides}.
With the decreasing value of the stability index $\alpha$ the probability of long jumps increases.
The action of long jumps is capable of slowing down the approach to the $\kappa\to\infty$ limit.
In Tab.~\ref{tab:splitting}, values of exponent $a$ fitted to the power-law dependence of $\pi_L$, see Eq.~(\ref{eq:pi-pl}), are provided.
Tab.~\ref{tab:splitting} shows that with the decreasing value of the stability index $\alpha$ the decay of the splitting probability slows down.
It demonstrates that strong noise pulses can weaken the overall role played by the drift.

\begin{table}
    \begin{tabular}{l || c | c | c }
     $\alpha$\;\; & \;\;$x=-0.5$\;\; & \;\;$x=0$\;\; & \;\;$x=0.5$\;\; \\ \hline \hline
    $0.5$ \;\;& 0.93 & 0.97 & 1.02 \\ \hline
    $0.75$ &  0.94 & 0.98 & 1.04\\ \hline
    $1$ & 0.98 & 1.02 & 1.05  \\ \hline
    $1.5$ & 1.13 & 1.15 & 1.18 \\
    \end{tabular}
    \caption{Values of the fitted exponents $a$ to the power-law decay of the splitting probability $\pi_L$, see Eq.~(\ref{eq:pi-pl}).}
    \label{tab:splitting}
\end{table}

In addition to the MFPT and splitting probability $\pi_L$, the escape kinetics can be characterized by the last hitting density $p_l(x)$, i.e., the distribution of last points visited prior to leaving the domain of motion.
The last hitting point density is an important characteristic of escape kinetics for processes with discontinuous trajectories, like L\'evy flights which are studied here.
Distribution of last hitting points for different $\alpha$ displays some common properties but also meaningful differences.
For large $\kappa$ the deterministic drift is significantly stronger than the stochastic part.
Therefore, the distribution of the last hitting point is narrowly concentrated near the boundary pointed by the drift.
When $\kappa$ is of the order of unity, the particle not only escapes via the boundary indicated by the drift but also starts escaping from the initial position, e.g., the middle of the interval.
Moreover, due to the strong drift, the particle most likely explores the space in the direction of the deterministic force.
With the further decrease of $\kappa$, noise becomes a dominating force in the system.
Consequently, the last hitting point density becomes more symmetric.
Also for $\kappa \ll 1$ the difference between different $\alpha$ is well visible.
It is not surprising, because, as it was mentioned above, this is the limit where the noise plays the dominating role.
However, even for $\kappa \simeq 1$, the differences between different values of stability indexes are also non-negligible.
For $\alpha<1$, particles move due to the action of the drift term, but escape mostly via long jumps.
For $\alpha>1$, noise allows particles to explore the whole interval, and therefore escapes, even though caused by a long jump, can be from the whole interval, c.f., blue dots and red triangles in the Fig.~\ref{fig:LHP}(b).

%
%
\section{Summary and conclusions \label{sec:summary}}

The noise-driven escape from a finite interval is a seminal, widely studied model of  stochastic dynamics.
It has been explored under the action of various noise types including the $\alpha$-stable L\'evy type noises.
Here, we have explored the competition between the deterministic drift and the stochastic, heavy-tailed driving.

The drift is an important factor which can facilitate the escape kinetics, as measured by the mean first passage time, from wide intervals by introducing a preferred direction of motion.
Importantly, the appropriately scaled MFPT follows the universal pattern as a function of the generalized P\'eclet number.
For large values of the generalized P\'ecklet number the drift is the dominating factor inducing the escape from a finite interval.
Nevertheless, the analysis of the splitting probability indicates that even if the drift dominates, escapes via a boundary opposite to the drift direction can occur.
The chance of `counter-drift' escapes increases with the decreasing value of the stability index $\alpha$ because with the decreasing $\alpha$ longer jumps become more probable, which is in accordance with the decomposition of $\alpha$-stable processes.
The probability of escaping via a boundary opposite to the drift direction decays as a power-law of the generalized P\'eclet number.
The exponent characterizing the decay of the splitting probability is the increasing function of $\alpha$.
For small values of the generalized P\'eclet number the drift is the main factor determining the escape kinetics.
Consequently, the splitting probability is sensitive to the initial condition.
Finally, for $\kappa \simeq 1$, the competition between deterministic drift and random force is recorded.

%
%
\section*{Acknowledgements}

This research was supported in part by PLGrid Infrastructure and by the National Science Center (Poland) grant 2018/31/N/ST2/00598.

\section*{Data availability}
The data (generated randomly using the model presented in the paper) that support the findings of this study are available from the corresponding author (KC) upon reasonable request.

%
%
\appendix
\section*{Appendices}

For completeness of the presentation, we recall basic information regarding the escape from finite intervals under the action of the Gaussian white noise and L\'evy white noise. For the latter case, the theoretical approach is more limited than for the former one.

\section{Gaussian white noise driving\label{sec:gwndriving}}

The overdamped Langevin equation takes the general form \cite{horsthemke1984,risken1996fokker}
\begin{equation}
  \frac{dx}{dt}=-\frac{dV(x)}{dx}+\sigma\xi(t),
  \label{eq:langevin-full}
\end{equation}
where $-V'(x)$ is the deterministic force and $\xi(t)$ is the Gaussian white noise satisfying $\langle \xi(t) \rangle=0$ and $\langle \xi(t)\xi(s) \rangle = \delta(t-s)$.
For the linear potential ($\mu>0$)
\begin{equation}
    V(x)=-\mu x
\end{equation}
Eq.~(\ref{eq:langevin-full}) attains the following form
\begin{equation}
  \frac{dx}{dt}=\mu+\sigma\xi(t).
  \label{eq:langevin-full2}
\end{equation}
Probability $p(x,t)=p(x,t|x_0,t_0)$ of finding the particle at time $t$ in the vicinity of $x$ evolves according to the (forward) Smoluchowski-Fokker-Planck equation \cite{goelrichter1974,gardiner2009}
\begin{equation}
    \frac{\partial p(x,t)}{\partial t} = \frac{\partial }{\partial x} \left[ -\mu p(x,t) + \frac{\sigma^2}{2} \frac{\partial p(x,t)}{\partial x} \right].
    \label{eq:forward-fp}
\end{equation}
The solution of Eq.~(\ref{eq:forward-fp}) can be found using the method of images or separation of variables \cite{cox1965}.
The MFPT $\mathcal{T}(x)$ from the $(-L,L)$ interval satisfies the backward Smoluchowski-Fokker-Planck equation \cite{goelrichter1974,gardiner2009}
\begin{equation}
    \mu \frac{\partial \mathcal{T}(x)}{\partial x} + \frac{\sigma^2}{2} \frac{\partial^2 \mathcal{T}(x)}{\partial x^2}=-1
    \label{eq:backward-fp}
\end{equation}
with the additional constraint $\mathcal{T}(\pm L)=0$.
The MFPT can be obtained from Eq.~(\ref{eq:backward-fp}) or from the general formula for the MFPT \cite[Eq.~(5.2.21)]{gardiner2009}
\begin{equation}
    \mathcal{T}(x)=\frac{L \left(e^{\frac{4 \mu  L}{\sigma ^2}}-2 e^{\frac{2 \mu  (L-x)}{\sigma
   ^2}}+1\right)-x e^{\frac{4 \mu  L}{\sigma ^2}}+x}{\mu  \left(e^{\frac{4 \mu
   L}{\sigma ^2}}-1\right)}.
   \label{eq:mfpt-full}
\end{equation}
From the general formula (\ref{eq:mfpt-full}) one can calculate the weak noise limit
\begin{equation}
    \lim_{\sigma\to 0} \mathcal{T}(x)= \frac{L-x}{\mu}
\end{equation}
and the weak drift limit
\begin{equation}
    \lim_{\mu\to 0} \mathcal{T}(x)= \frac{L^2-x^2}{\sigma^2}.
    \label{eq:mfpt-free}
\end{equation}
For a weak diffusion (small $\sigma$) with $\mu>0$ the particle deterministically moves along the drift towards the (right) absorbing boundary and the MFPT is equal to the distance divided by the drift strength (``velocity'').
For the weak drift, the motion becomes purely diffusive and the MFPT is equal to the MFPT of a free particle from the finite interval restricted by two absorbing boundaries.
Analogously, for the strong diffusion and the strong drift one obtains
\begin{equation}
    \lim_{\sigma\to \infty} \mathcal{T}(x)=0
\end{equation}
and
\begin{equation}
    \lim_{\mu\to \infty} \mathcal{T}(x)=0,
\end{equation}
because the particle immediately leaves the domain of motion.

The probability $\pi_L(x)$ of leaving the $(-L,L)$ interval through the left boundary satisfies \cite{goelrichter1974,gardiner2009}
\begin{equation}
\mu \frac{\partial \pi_L(x)}{\partial x}     +\frac{\sigma^2}{2} \frac{\partial^2 \pi_L(x)}{\partial x^2} = 0.
\end{equation}
with the boundary condition $\pi_L(-L)=1$ and $\pi_L(L)=0$.
The splitting probability is given by
\begin{equation}
    \pi_L(x)=\frac{e^{\frac{2 \mu  (L-x)}{\sigma ^2}}-1}{e^{\frac{4 \mu  L}{\sigma ^2}}-1}
    \label{eq:splitting-full}
    \end{equation}
It can be also obtained from  \cite[Eq.~(5.2.128)]{gardiner2009}.
The probability $\pi_R(x)$   to escape through the right absorbing boundary reads
\begin{equation}
    \pi_R(x)=1-\pi_L(x).
\end{equation}

From the general formula (\ref{eq:splitting-full}) one can calculate the weak noise limit
\begin{equation}
    \lim_{\sigma\to 0} \pi_L(x)= 0
\end{equation}
and the weak drift limit
\begin{equation}
    \lim_{\mu\to 0} \pi_L(x)= \frac{L-x}{2L}.
\end{equation}
For the weak diffusion (small $\sigma$) with $\mu>0$ the particle deterministically moves along the drift toward the right absorbing boundary, therefore the probability to escape through the left boundary vanishes.
For weak drift, the motion becomes purely diffusive and the splitting probability is equal to the $\pi_L(x)$ of a free particle from the finite interval restricted by two absorbing boundaries.
Analogously, for the strong diffusion and the strong drift one obtains
\begin{equation}
    \lim_{\sigma\to \infty} \pi_L(x)=\frac{L-x}{2L}
\end{equation}
and
\begin{equation}
    \lim_{\mu\to \infty} \pi_L(x)=0,
\end{equation}
because the motion is insensitive to the drift ($\sigma\to\infty$) or the particle deterministically moves to the right ($\mu\to\infty$).

Finally,  for $x=0$, the formula (\ref{eq:mfpt-full}) simplifies to
\begin{equation}
    \mathcal{T}(0)=\frac{L}{\mu} \tanh\left[ \frac{L\mu}{\sigma^2} \right].
\end{equation}
and Eq.~(\ref{eq:splitting-full}) assumes the form
\begin{equation}
    \pi_L(0)=\frac{1}{e^{\frac{2 \mu  L}{\sigma ^2}}+1}.
\end{equation}

\section{L\'evy white noise driving\label{sec:lwndriving}}

For $\alpha$-stable driving in Eq.~(\ref{eq:langevin-full}) the GWN is replaced by L\'evy noise.
For $\mu=0$ and the symmetric $\alpha$-stable noise, see Eq.~(\ref{eq:fcharakt}), the MFPT is given by \cite{getoor1961,widom1961stable,kesten1961random,kesten1961theorem,zoia2007}
\begin{equation}
    \mathcal{T}(x)=\frac{(L^2-x^2)^{\alpha/2}}{\Gamma(1+\alpha)\sigma^\alpha}.
    \label{eq:mfpt-alpha-app}
\end{equation}
$\alpha$-stable noise with $\alpha=2$ is equivalent to the Gaussian white noise.
The formula (\ref{eq:mfpt-alpha-app}) with $\alpha=2$ differs from Eq.~(\ref{eq:mfpt-free}) by the factor 2 in the denominator.
This factor arises because $\alpha$-stable density with $\alpha=2$ is equivalent to the Gaussian distribution with the standard deviation equal to $\sqrt{2}$, i.e., $N(0,\sqrt{2})$.
Consequently, for the $\alpha$-stable driving the particle escapes two times faster than for the Gaussian white noise with the same scale parameter $\sigma$.

Information on further properties of escape kinetics can be found in a series of earlier works, e.g., first hitting points \cite{blumenthal1961}, leapovers \cite{ray1958stable,koren2007}, first arrivals \cite{blumenthal1961}, and the mini review \cite{dybiec2016jpa}.

%
%


\def\url#1{}

\end{document}